\documentclass[journal,twocolumn,twoside,times,10pt]{IEEEtran}
\usepackage{epsfig,amsfonts,subfigure,color}
\usepackage{graphicx,cite,amssymb,amsmath,mathrsfs}
\newtheorem{theorem}{\indent Theorem}[section]

\newcommand{\D}{\mathrm{d}}

\renewcommand{\IEEEQED}{\IEEEQEDopen}

\begin{document}


\title{Graph-Based Random Access for the Collision Channel without Feedback: Capacity Bound}
\author{
Enrico Paolini, Gianluigi Liva and Marco Chiani
\thanks{Enrico
Paolini and Marco Chiani are with DEIS/WiLAB, University of Bologna,
via Venezia 52, 47521 Cesena (FC), Italy.
Email:\texttt{\{e.paolini,marco.chiani\}@unibo.it.}}
\thanks{Gianluigi Liva is with the Institute of Communications
and Navigation, German Aerospace Center (DLR), Oberpfaffenhofen, 82234
Wessling, Germany. Email: \texttt{Gianluigi.Liva@dlr.de}.}
\thanks{Supported in part by the EC under Project FP7 OPTIMIX (ICT-214625).}}
\maketitle
\thispagestyle{empty} \pagestyle{empty}


\begin{abstract}
A random access scheme for the collision channel
without feedback is proposed. The scheme is based on erasure
correcting codes for the recovery of packet segments that are lost
in collisions, and on successive interference cancellation for
resolving collisions. The proposed protocol achieves reliable
communication in the asymptotic setting and attains capacities close
to $1\,\mbox{\boldmath $\mathrm{[packets/slot]}$}$. A capacity bound as a
function of the overall rate of the scheme is
derived, and code distributions tightly approaching the bound developed.
\end{abstract}

{\pagestyle{plain} \pagenumbering{arabic}}


\section{Introduction}\label{sec:intro}

Since the introduction of the ALOHA protocol \cite{Abramson:ALOHA},
several random access (RA) {schemes} have been introduced. Among
them, some feedback-free RA protocols originally proposed in
\cite{Massey85:collision_channel,Hui84:CCwoFB} re-gained attention
in the recent past
\cite{Thomas00:Capacity_Wireless_Collision,Grant05:CCwithRecovery,Shum09:Shift_Invariant_Protocol}.
In \cite{Massey85:collision_channel}, the capacity of the so-called
collision channel without feed-back (CCw/oFB) was analyzed, assuming
slot-aligned but completely asynchronous users' transmissions.
Moreover, a simple approach to achieve error-free transmission over
the CCw/oFB was proposed. The approach of
\cite{Massey85:collision_channel} consists of assigning different
periodic protocol (access) sequences to the users. Each sequence
defines in which slots each user is allowed to access the shared
channel. Furthermore, the users encode their packets by means of
erasure correcting codes. The user's packet can be recovered
whenever a sufficient number of codeword segments are received
collision free. Hence, by selecting proper protocol sequences, it is
possible to ensure that a sufficient number of segments per user are
recovered, even if the beginning of the different protocol sequences
is unsynchronized. In this way, a capacity equal to
$1/e\,\mathrm{[packets/slot]}$ is achieved as $M\rightarrow
\infty${,} where $M$ is the number of users accessing the RA
channel. The same capacity is achieved also in the unslotted case.
Although simple, the approach of \cite{Massey85:collision_channel}
poses some challenges, especially if a large (and varying) number of
users has to be served
\cite{Hui84:CCwoFB,Thomas00:Capacity_Wireless_Collision}.

Recently, RA schemes profiting from SIC have been introduced and
analyzed
\cite{DeGaudenzi07:CRDSA,Giannakis07:SICTA,Liva11:IRSA,Paolini11:CSA_ICC}.
These schemes share the feature of canceling the interference caused
by collided packets on the slots where they have been transmitted
whenever a clean (uncollided) copy of them is detected. In
\cite{Liva11:IRSA,Paolini11:CSA_ICC} it was shown that the
successive interference cancellation (SIC) process can be well
modeled by means of a bipartite graph. The analysis proposed in
\cite{Liva11:IRSA,Paolini11:CSA_ICC} resembles {density evolution}
analysis of low-density parity-check (LDPC) and doubly-generalized
LDPC (D-GLDPC) codes over erasure channels
\cite{studio3:GallagerBook,rich01:design,paolini10:random}. By
exploiting design techniques from the LDPC context, a
remarkably-high throughput (e.g. up to
$0.8\,\mathrm{[packets/slot]}$) can be achieved in practical
implementations. The schemes considered in
\cite{DeGaudenzi07:CRDSA,Giannakis07:SICTA,Liva11:IRSA,Paolini11:CSA_ICC}
assume a feedback from the receiver to achieve a zero packet loss
rate. Furthermore, the approaches proposed in
\cite{DeGaudenzi07:CRDSA,Liva11:IRSA} has been recently considered
for inclusion in the next generation Digital Video Broadcasting
Return Channel via Satellite (DVB-RCS) standard \cite{RCSNG}.

In this paper, we investigate the application of SIC to collision
channels without feedback. With respect to
\cite{Massey85:collision_channel}, the proposed RA scheme does
\emph{not} require the assignment of protocol sequences to users,
who select time slots for their transmissions in a randomized and
uncoordinated fashion. The proposed scheme stems from the coded
slotted ALOHA (CSA) protocol proposed in \cite{Paolini11:CSA_ICC}.

Each user splits his generic packet into segments, that are encoded
through a binary linear (erasure) block code chosen on a
packet-by-packet basis, according to some probability distribution.
The resulting encoded segments are then transmitted in time slots
selected with uniform probability out of the slots composing the
medium access control (MAC) frame. If for a packet a sufficient
number of encoded segments are correctly received, its entire set of
encoded segments can be recovered by the erasure code. Moreover, the
contribution of interference of cleaned segments can be canceled
from the slots in which they caused collisions. It is shown how,
iterating this process, for a channel traffic below a threshold
value, arbitrarily low error probabilities can be achieved. We refer
to this threshold as the \emph{capacity} of the scheme. By suitably
selecting the distribution of erasure codes, capacities close to
$1\,\mathrm{[packets/slot]}$ can be obtained. It is also shown that
there is a fundamental trade-off between the rate of the scheme and
its capacity. An upper bound on the capacity is proved, and code
distributions tightly approaching it derived.


\section{System Model}\label{sec:model}
We consider {a} slotted RA scheme where slots are grouped in MAC
frames, all with the same length (in slots). Each user is frame- and
slot-synchronous, and attempts one \emph{burst} (i.e., packet)
transmission per MAC frame. Consider $M$ users, each attempting the
transmission of a burst of time duration $T_{\mathrm{slot}}$ over a
MAC frame of time duration $T_{\mathrm{frame}}$. Neglecting guard
times, the MAC frame is composed of
$N=T_{\mathrm{frame}}/T_{\mathrm{slot}}$ slots. We say that the
offered (channel) traffic is $G=M/N$.

\begin{figure}[t]
\begin{center}
\includegraphics[width=0.8\columnwidth,draft=false]{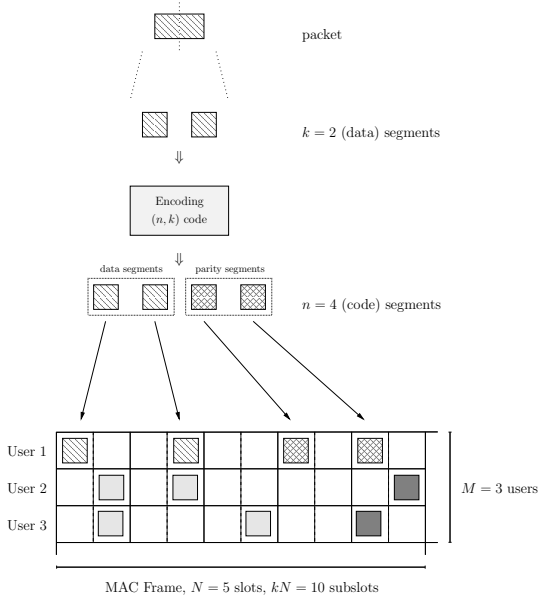}
\end{center}
\caption{Model of the access scheme. Each user splits his bursts into $k=2$ segments. User $1$
encodes them by a $(4,2)$ linear block code, users $2$ and $3$ by a $(3,2)$ single parity-check
(SPC) code. The darkened squares denote parity segments.}\label{fig:CRA_Model}
\end{figure}

The proposed access scheme works as follows. Prior to transmission
of a user's burst of time duration $T_{\mathrm{slot}}$, the burst is
divided into $k$ information \emph{segments}, each of time duration
$T_{\mathrm{seg}}=T_{\mathrm{slot}}/k$. The $k$ information segments
are then encoded by the user via a packet-oriented linear block code
generating $n_h$ encoded segments, each of time duration
$T_{\mathrm{seg}}$. For each transmission, the $(n_h,k)$ code is
chosen randomly by the user from a set
$\mathcal{C}=\{\mathscr{C}_1,\mathscr{C}_2,\dots,\mathscr{C}_{n_c}\}$
of $n_c$ candidate codes. For $h\in\{1,2,\dots,n_c\}$ the code
$\mathscr{C}_h$ has length $n_h$, dimension $k$, rate $R_h=k/n_h$,
and minimum distance at least $2$. Each code $\mathscr{C}_h$ has no
idle symbols. At any transmission, each user draws his local code
from the set $\mathcal{C}$ independently of his previous choices and
without any coordination with the other users. The code is drawn
according to a probability mass function (p.m.f.) {\boldmath
$\Lambda$}$=\{\Lambda_h\}_{h=1}^{n_c}$ which is the same for all
users. The MAC frame is composed of $kN$
\mbox{(sub-)}slots.\footnote{The definition of MAC frame as sequence
of $N$ slots is instrumental to the definition of load $G$ only. The
actual minimum unit that can be allocated to a segment transmission
are the sub-slots.} The $n_h$ coded segments are then transmitted by
the user over $n_h$ slots picked uniformly at random. An example for
$k=2$ is provided in Fig.~\ref{fig:CRA_Model}. For the special case
of $k=1$, $T_{\mathrm{seg}}=T_{\mathrm{slot}}$ and each
$\mathscr{C}_h$ is a repetition code of length $n_h$. The overall
\emph{rate} of the scheme is defined as $R=k/\bar{n}$, where
$\bar{n} := \sum_{h=1}^{n_c} {\Lambda_h} n_h$ is the expected length
of the code.\footnote{Note that $R$ does not represent the average
code rate {$\bar{R}$} adopted by {each user}. It is easy to prove
(Jensen's inequality) that the average rate $\bar{R}=\sum_h
\Lambda_h R_h$ is lower {bounded} by $R$.} {We also define the
polynomial $\Lambda(x)=\sum_{h=1}^{n_c} \Lambda_h x^h$ and the
polynomial $\lambda(x)=\sum_{h=1}^{n_c} \lambda_h x^{h-1}$, where
$\lambda_h=\Lambda_h n_h / (\sum_{i=1}^{n_c} \Lambda_i n_i)$. We
refer to $\Lambda(x)$ as the component code distribution.}

\begin{figure}[t]
\begin{center}
\includegraphics[width=0.7\columnwidth,draft=false]{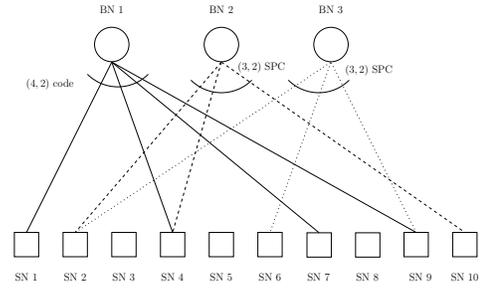}
\end{center}
\caption{Bipartite graph representation of the access scheme of Fig.~\ref{fig:CRA_Model}. Subslots
are represented by SNs, bursts by BNs. Each segment corresoends to an
edge.}\label{fig:CRA_Graph}
\end{figure}

We adopt a graph representation of the RA scheme, depicted in
Fig.~\ref{fig:CRA_Graph} (related to the same example as
Fig.~\ref{fig:CRA_Model}). Consider $M$ users transmitting over a
frame composed of $kN$ slots. The situation can be represented by a
bipartite graph, $\mathcal{G} =
(\mathcal{B},\mathcal{S},\mathcal{E})$, consisting of a set
$\mathcal{B}$ of $M$ \emph{burst nodes} (one for each user), a set
$\mathcal{S}$ of $kN$ \emph{sum nodes} (one for each slot), and a
set $\mathcal{E}$ of edges. An edge connects the $i$-th burst node
(BN) to the $j$-th sum node (SN) if and only if an encoded segment
associated with the burst of the $i$-th user is transmitted in the
$j$-th slot. The number of edges connected to a BN or SN is the node
degree. Therefore, a burst encoded via the code $\mathscr{C}_h$ is
represented as a degree-$n_h$ BN, and a slot where $d$ segments
collide as a degree-$d$ SN. Moreover, a BN where $\mathscr{C}_h$ is
used during the current transmission is named a BN of type~$h$.

Each coded segment associated with a type-$h$ BN is equipped with
information about its relevant user and with pointers to the other
$n_h-1$ segments of the user.\footnote{In practical implementations,
the overhead due to the inclusion of pointers in the segment header
may be reduced by adopting more efficient techniques. For fixed $k$,
one may include in the segment header the code index $h$ together
with a random seed, out of which it is possible to reconstruct (by a
pre-defined pseudo-random number generator) the positions of the
$n_h$ segments.} We assume that collisions are always detected at
the receiver and that they determine a complete loss of information
about the colliding segments. We also assume that, when a burst is
received in a clean slot, it is always successfully decoded.
Therefore, after transmission, we can think of each BN as connected
to ``known'' edges and to ``unknown'' ones, so that some of its
information segments are known, and the others unknown. At a BN of
type $h$, erasure decoding of the code $\mathscr{C}_h$ may recover
some of the unknown encoded and information segments. It is now
possible to subtract the interference contribution of the newly
recovered encoded segments from the signal received in the
corresponding slots. If $d-1$ segments that collided in a SN of
degree $d$ have been recovered by the corresponding BNs, the
remaining segment becomes known. The SIC process combined with local
decoding at the BNs proceeds iteratively, i.e., cleaned segments may
allow to resolve other collisions. In this paper, ideal SIC is
assumed.\footnote{To summarize, our analysis relies on three
assumptions: i) \emph{Destructive (and detectable) collisions}; ii)
\emph{Sufficiently high signal-to-noise ratio (SNR)} (segments
received in clean slots are always successfully decoded); iii)
\emph{Ideal channel estimation} (an hypothesis under which, together
with the previous one, we can assume an ideal SIC). These
assumptions simplify the analysis without substantially affecting
the performance, as shown in \cite{DeGaudenzi07:CRDSA,Liva11:IRSA}.}
This procedure is equivalent to iterative decoding of a D-GLDPC code
over the erasure channel, where variable nodes are generic linear
block codes and check nodes are single-parity-check (SPC) codes.

\section{Asymptotic Analysis}\label{sec:asymptotic_analysis}
In this section, the evolution of the SIC process for given $k$ and
$G$, in the asymptotic case where $M \rightarrow \infty$ (and $N
\rightarrow \infty$) is {overviewed}. {Most of} the results
presented in this section were developed in
\cite{Paolini11:CSA_ICC}. This serves to define the extrinsic
information transfer (EXIT) functions for the BN and SN sets, used
in the proof of {our} main result {presented in
Section~\ref{sec:capacity}}, and to formalize the concept of
{capacity} under iterative SIC.

We start by briefly recalling the definition of information function of a linear block code
\cite{helleseth97:information}.
Consider an $(n,k)$ linear block code $\mathscr{C}$, where $n$ is the codeword length and $k$ the
code dimension, and let $\mathbf{G}$ be any generator matrix of $\mathscr{C}$. Then, the $g$-th
un-normalized information function of $\mathscr{C}$, denoted by $\tilde{e}_g$, is defined as the
summation of the ranks {of all} possible submatrices obtained selecting $g$ columns (with $0
\leq g \leq n$) out of $\mathbf{G}${, regardless their order}.

Next, assume that maximum a-posteriori (MAP) decoding is used
locally at each BN. At the $i$-th iteration of the {S}IC process,
let $p_{i-1}$ be the average probability that an edge is connected
to a SN associated with a slot where a collision still persists,
before MAP decoding is performed at each BN. Moreover, let $q_i$ be
the average probability that an edge is connected to a BN whose
contribution of interference on the corresponding SN cannot be yet
canceled, after MAP decoding has been performed at each BN. Then:
\begin{align}\label{eq:q_i(p_i-1)}
q_i & = \frac{1}{\bar{n}}\sum_{h=1}^{n_c} \Lambda_h \sum_{t=0}^{n_h-1} p_{i-1}^t
(1-p_{i-1})^{n_h-1-t} [(n_h-t) \tilde{e}^{(h)}_{n_h-t} \notag\\
& \phantom{=} - (t+1) \tilde{e}^{(h)}_{n_h-1-t}] =: f_{\mathsf{b}}(p_{i-1}) \, .
\end{align}
Adopting a {consolidated} terminology in the theory of modern error
correcting codes, the function $f_{\mathsf{b}}(p)$ is referred to as
the average EXIT function of the BN set. It is easy to verify
that %
$
f_{\mathsf{b}}(p) = \sum_{h=1}^{n_c} \lambda_h f_{\mathsf{b}}^{(h)}(p)\, ,
$ %
where
\begin{align}
f_{\mathsf{b}}^{(h)}(p) := & \,\frac{1}{n_h}\sum_{t=0}^{n_h-1} p^t (1-p)^{n_h-1-t}
[(n_h-t) \tilde{e}^{(h)}_{n_h-t} \notag \\
&\, - (t+1) \tilde{e}^{(h)}_{n_h-1-t}]
\end{align}
is called the average EXIT function of a type-$h$ BN, under MAP
decoding.

Equation \ref{eq:q_i(p_i-1)} allows to update $q_i$ given $p_{i-1}$.
The dependence of $p_i$ on $q_i$ is instead obtained by observing
that the number of segments received in a slot follows the
distribution
\[
\Psi_l={M\choose l}
\left(\frac{\bar{n}G}{kM}\right)^l\left(1-\frac{\bar{n}G}{kM}\right)^{M-l}\, ,
\]
{where $\Psi_l$ is the probability to receive $l$ segments in the generic slot.} {D}efining
$\Psi(x) =\sum_l \Psi_l x^l$ and letting $M\rightarrow \infty$, {yields}
\[
\Psi(x)=\exp\left(-\frac{G}{R}(1-x)\right).
\]
It follows that $p_i$ can be obtained from $q_i$ by the relationship
\begin{align}\label{eq:p_i(q_i)}
p_i = 1-\rho(1-q_i) = 1 - e^{-\frac{G}{R}q_i } =: f_{\mathsf{s}}(q_i)\, ,
\end{align}
{where} $\rho(x)=\Psi'(x)/\Psi'(1)$ and $\Psi'(x)=\mathrm{d}
\Psi(x)/\mathrm{d}x$. The function $f_{\mathsf{s}}(q)$ is called the
average EXIT function of the SN set. The equations
\eqref{eq:q_i(p_i-1)} and \eqref{eq:p_i(q_i)} define a discrete-time
dynamical system $q_i=q_i(q_{i-1})$ with starting point
$q_1=f_{\mathsf{b}}(0)$, whose stability was analyzed in
\cite{Paolini11:CSA_ICC}. Note that the normalized offered traffic
$G$ is involved in the recursion through \eqref{eq:p_i(q_i)}. The
asymptotic threshold of the SIC process, denoted by
$G^*=G^*(\mathcal{C},\mbox{\boldmath $\Lambda$})$, is defined as the
sup of the ensemble of all $G\geq 0$ such that $q_i\rightarrow 0$ as
$i\rightarrow\infty$. In the asymptotic setting
$M\rightarrow\infty$, for all $G<G^*(\mathcal{C},\mbox{\boldmath
$\Lambda$})$ the throughput is $S=G$, i.e., all collisions are
resolved. In this sense, $G^*(\mathcal{C},\mbox{\boldmath
$\Lambda$})$ represents the \emph{capacity} of the RA scheme
conditional to the specific choice of
$\mathcal{C}=\{\mathscr{C}_1,\mathscr{C}_2,\dots,\mathscr{C}_{n_c}\}$
and \mbox{\boldmath $\Lambda$}.\footnote{Readers with a background
in modern coding theory may prefer the nomenclature ``asymptotic
threshold'' to ``capacity''.}

\begin{figure}[t]
\begin{center}
\includegraphics[width=0.90\columnwidth,draft=false]{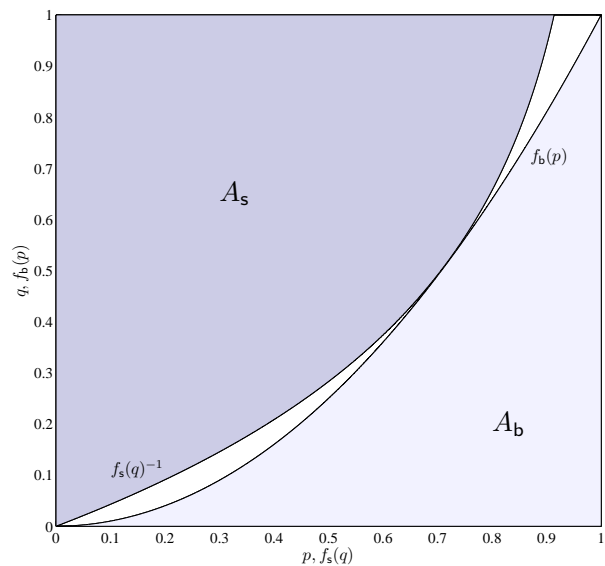}
\end{center}
\caption{EXIT chart for a regular coded random access scheme employing a
rate-$1/3$ repetition code at each BN, characterized by $G^*=0.816$.
}\label{fig:EXIT}
\end{figure}

The recursion defined by \eqref{eq:q_i(p_i-1)} and
\eqref{eq:p_i(q_i)} can be visualized in a so-called \emph{EXIT
chart}, which displays  $f_{\mathsf{b}}(p)$ vs.
$f_{\mathsf{s}}^{-1}(p)$. An example of EXIT chart for a scheme
employing a $(3,1)$ repetition code at each BN is provided in
Fig.~\ref{fig:EXIT}. As the iteration index $i$ increases, the
evolution of the pair of probabilities $(p_i,q_i)$ traces a zig-zag
pattern inside the tunnel between the two curves. Whenever we
operate the scheme below its capacity,
$G<G^*(\mathcal{C},\mbox{\boldmath $\Lambda$})$, the two curves do
not intersect, leaving the tunnel open. This lets the probabilities
$(p,q)$ achieve the $(0,0)$ point. On the contrary, if the scheme is
operated above its capacity, $G>G^*(\mathcal{C},\mbox{\boldmath
$\Lambda$})$, the two curves intersect (closing the tunnel) in a
point $(\hat{p},\hat{q})$ with $\hat{p}>0, \hat{q}>0$, and the
iterative SIC process converges to a fixed point corresponding to a
non-zero residual erasure probability.

\section{Capacity Bound}\label{sec:capacity}

In this section, we present our main result along with its proof.
The proposed proof is based on the EXIT functions reviewed in
Section~\ref{sec:asymptotic_analysis} {and on the Area Theorem, a
well-known result in the theory of EXIT functions
\cite{Ashikhmin:AreaTheorem}.} An alternative proof (not proposed
here for space reasons) can be given exploiting algebraic arguments.

\medskip
\begin{theorem}\label{theorem:capacity_bound}
For rational $R$ and $0<R \leq 1$, let $\bar{G}{(R)}$ be the unique positive solution of the
equation
\begin{align}\label{eq:barG_eqn}
G=1-e^{-G/R}
\end{align}
in $[0,1)$. Then, the capacity $G^*(\mathcal{C},\mbox{\boldmath
$\Lambda$})$ of the RA scheme fulfills
\begin{align}\label{eq:capacity_bound}
G^*(\mathcal{C},\mbox{\boldmath $\Lambda$}) < \bar{G}{(R)}
\end{align}
for \emph{any} choice of $\mathcal{C}=\{\mathscr{C}_1,\mathscr{C}_2,\dots,\mathscr{C}_{n_c}\}$ and
\mbox{\boldmath $\Lambda$} corresponding to a rate $R$.
\end{theorem}
\begin{IEEEproof} For given $\mathcal{C}$ and \mbox{\boldmath $\Lambda$}, the evolution of the
probabilities $(p_i,q_i)$ is governed by the recursion
$p_i=f_{\mathsf{s}}(q_{i})$ and $q_i=f_{\mathsf{b}}(p_{i-1})$ in
\eqref{eq:p_i(q_i)} and \eqref{eq:q_i(p_i-1)}, for all $i\geq 1$ and
with $q_1=f_{\mathsf{b}}(0)$. Let's furthermore denote the areas
below the BN and the SN EXIT functions (over the interval $[0,1]$)
by
\begin{align*}
A_{\mathsf{b}}=\int_0^1 f_{\mathsf{b}}(p)\D p \quad \mathrm{and} \quad
A_{\mathsf{s}}=\int_0^1 f_{\mathsf{s}}(q)\D q
\end{align*}
respectively. A necessary condition for successful decoding is to
have an open tunnel  between the two curves in the EXIT chart (see
Fig.~\ref{fig:EXIT}). For all $G\leq G^*(\mathcal{C},\mbox{\boldmath
$\Lambda$})$, we must have an open tunnel in the EXIT chart, which
implies
\begin{equation}\label{eq:AREA_CONDITION}
A_{\mathsf{b}}+A_{\mathsf{s}}<1\, .
\end{equation}
In particular, \eqref{eq:AREA_CONDITION} must be satisfied for
$G=G^*(\mathcal{C},\mbox{\boldmath $\Lambda$})$. The area below the
SN EXIT function \eqref{eq:p_i(q_i)} is given by
\begin{equation}
A_{\mathsf{s}}=1+\frac{R}{G} e^{-\frac{G}{R}}-\frac{R}{G}\, .\label{eq:AREA_S}
\end{equation}
Moreover, the area below the BN EXIT function \eqref{eq:q_i(p_i-1)}
is given~by
\begin{align}\label{eq:AREA_B}
A_{\mathsf{b}}&=\sum_{h=1}^{n_c} \lambda_h \int_0^1 f^{(h)}_{\mathsf{b}}(p) \D p = \sum_{h=1}^{n_c}
\lambda_h \frac{k}{n_h},
\end{align}
where the second equality holds under the assumption of MAP erasure
decoding at the burst node, and is due to the Area Theorem {in}
\cite{Ashikhmin:AreaTheorem}, which states that the area below the
MAP EXIT function of a binary linear block code without idle symbols
equals its code rate. By incorporating \eqref{eq:AREA_S} and
\eqref{eq:AREA_B} in \eqref{eq:AREA_CONDITION} we {obtain}
\begin{equation}
\sum_{h=1}^{n_c}
\lambda_h\frac{k}{n_h}+1+\frac{R}{G}
e^{-\frac{G}{R}}-\frac{R}{G}<1.\label{eq:AREA_CONDITION_EXPANDED}
\end{equation}
Observing that $\lambda_h=n_h\Lambda_h/\sum_l (n_l
\Lambda_l)=n_h\Lambda_h/\bar{n}$, we have $\sum_h k\lambda_h /
n_h=k\sum_h \Lambda_h/\bar{n} = k/\bar{n}=R$. Thus,
\eqref{eq:AREA_CONDITION_EXPANDED} reduces~to
$R+(R/G) e^{-G/R}<R/G$
which, for $G=G^*(\mathcal{C},\mbox{\boldmath $\Lambda$})$ yields
\begin{equation}\label{eq:AREA_CONDITION_FINAL}
G^*(\mathcal{C},\mbox{\boldmath $\Lambda$})<1- e^{-\frac{G^*(\mathcal{C},{\scriptsize
\mbox{\boldmath $\Lambda$}})}{R}}\,.
\end{equation}
Defining $\bar{G}(R)$ as the unique solution in $(0,1]$ of \eqref{eq:barG_eqn}, we have
$R(\bar{G})=-\bar{G}/\log(1-\bar{G})$. Moreover, from \eqref{eq:AREA_CONDITION_FINAL} we have
$R(G^*)>-G^*/\log(1-G^*)$. Since the function $y=-x/\log(1-x)$, $x\in[0,1)$, is monotonically
decreasing, we obtain \eqref{eq:capacity_bound}.
\end{IEEEproof}


\medskip
Note that, while $G^*(\mathcal{C},\mbox{\boldmath $\Lambda$})$ depends on $R$ through $\mathcal{C}$
and \mbox{\boldmath $\Lambda$}, its upper bound $\bar{G}(R)$ depends \emph{solely} on $R$.

\section{Capacity-Approaching Schemes}\label{sec:capacity_distributions}
Theorem~\ref{theorem:capacity_bound} establishes an upper bound on
the capacity of the proposed RA scheme. In this section, we show
that this bound can be indeed approached. To do so, for a given $k$,
a given set
$\mathcal{C}=\{\mathscr{C}_1,\mathscr{C}_2,\dots,\mathscr{C}_{n_c}\}$
of component codes, and a given target rate $R$, we generated by
differential evolution (DE) \cite{studio3:storn1997} optimization
the distribution \mbox{\boldmath $\Lambda$} which maximizes
$G^*(\mathcal{C},\mbox{\boldmath $\Lambda$})$. In order  to limit
the search space, we focused on schemes based on codes of
moderate-low length.

\begin{figure}[t]
\begin{center}
\includegraphics[width=0.84\columnwidth,draft=false]{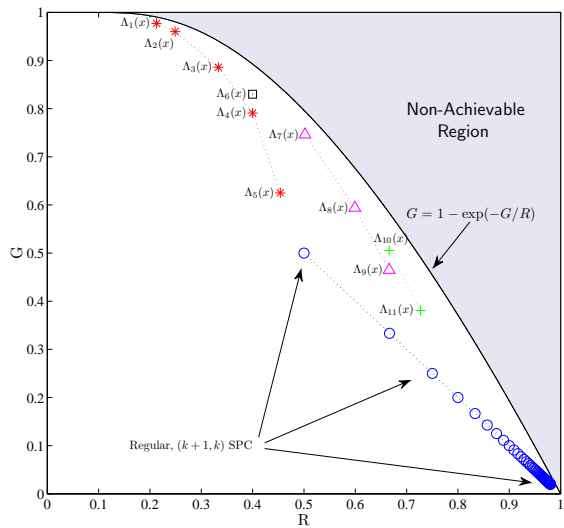}
\end{center}
\caption{Upper bound to the capacity vs. rate $R$. Thresholds $G^*$ are reported for
selected distributions. Distributions $\Lambda_i(x), i=1,\ldots, 5$  (\textcolor{red}{$\ast$}) are
based on repetition codes ($k=1$). Distribution $\Lambda_6(x)$ ({$\square$}) is based on
MDS codes ($k=2$). Distributions $\Lambda_i(x), i=7,8,9$
(\textcolor{magenta}{$\vartriangle$}) are based on MDS codes ($k=3$). Distributions
$\Lambda_i(x), i=10,11$  (\textcolor{green}{$+$}) are based on MDS codes ($k=4$).
Distributions based on $(k+1,k)$ SPC codes are also displayed
(\textcolor{blue}{$\circ$}).}\label{fig:Capacity}
\end{figure}

For example, for the case of $k=1$ and $R=0.2$, we limited $n_h$ to $30$
(i.e., repetition codes with rate down to $1/30$ have been
considered). In this case, we obtained the distribution
\begin{align*}
&\Lambda_1(x)=0.494155x^2 +  0.159085x^3 +  0.107372x^4\\
&\!\!+ 0.070336x^5+ 0.045493x^6 +  0.019898x^7 + 0.024098x^{11}\\
&\!\! +  0.008636x^{12} +  0.005940x^{13} +  0.008749x^{15}  +  0.002225x^{18}\\
&\!\! +  0.001261x^{20} + 0.002607x^{22} +  0.008092x^{23} +  0.002287x^{24}\\
&\!\! + 0.012274x^{25} +  0.002530x^{26} + 0.003094x^{27} + 0.002558x^{28}\\
&\!\! +  0.005891x^{29} + 0.013419x^{30},
\end{align*}
where $\Lambda_1(x)=\sum_h \Lambda_h x^h$ and $\mathscr{C}_h$ is the
$(h,1)$ repetition code. The associated threshold is $G^*=0.977$.
The corresponding point on the $G\times R$ plane is reported in
Fig.~\ref{fig:Capacity} and compared with the bound given by
Theorem 1. On the same plot, other distributions, denoted by $\Lambda_i(x)$ for
$i\in\{2,\dots, 5\}$, are reported for $k=1$. Whereas for low
rates $R$ repetition-based scheme approach tightly the bound,
for rates close to $1/2$ they show visible losses. The
distribution $\Lambda_5(x)=0.8x^2 +  0.2x^3$ (obtained by allowing a
maximum $n_h$ of $5$) presents a rate $R=5/11\simeq 0.45$ and attains a
threshold $G^*=0.625$, whereas $\bar{G}(5/11)\simeq 0.843$. This effect if somehow expected, since
in
the limiting case of $R=1/2$ each user employs a $(2,1)$
repetition code, and the corresponding threshold is $G^*=0.5$.\footnote{For $k=1$,
$R=1/2$ represents a limiting case because, for rates $R>1/2$, the
rate-$1$ repetition code (i.e., no
coding) must be included in the set $\mathcal{C}$. In this case, it is easy to prove that the
threshold $G^*$ would drop to $0$.}

As the rate $R$ grows, it is convenient to adopt codes with $k>1$.
To this purpose, we designed further schemes where each
$\mathscr{C}_h\in\mathcal{C}$ is maximum distance separable
(MDS).\footnote{Imposing limits on $n_h$, this assumption is
realistic. For instance, (generalized) Reed-Solomon codes on finite
fields of moderate order may be used.} It is assumed that each burst
node adopts a sub-optimum (in the symbol-wise MAP sense) decoding
approach, which consists of decoding its lost segments only if the
number of its collision-free received segments is at least $k$.
Under this assumption, the EXIT function of a BN using an $(n_h,k)$
MDS codes is given~by
\begin{align}
f_{\mathsf{b}}^{(h)}(p) =\sum_{l=0}^{k-1} {n_h-1 \choose l} (1-p)^{l}p^{n_h-l-1}.
\end{align}
We define in this case the distribution $\Lambda(x)=\sum_h \Lambda_h
x^h$, where $\mathscr{C}_h$ is an $(h+k,k)$ MDS code.

For $k=2$ we designed a distribution $\Lambda_6(x)=0.276023x +
0.366641x^2 + 0.127979x^3 + 0.229357x^7$ characterized by $R=0.4$
and $G^*=0.83$. For the same rate, the best found repetition-based
($k=1$) distribution for a maximum $n_h$ set to $10$, denoted by
$\Lambda_4(x)$, achieves $G^*=0.79$. Moving to the moderate-rate
regime, we heuristically found that the bound can be better
approached by adopting distributions based on higher code dimensions
(see Fig.~\ref{fig:Capacity}). We conjecture that the bound can be
approached even more tightly for larger code dimensions ($k$). In
Fig.~\ref{fig:Capacity} the thresholds achieved by regular schemes
based on SPC codes of increasing rates are provided. In has been
proved in \cite{Paolini11:CSA_ICC} that for these schemes the
threshold admits the close form $G^*=(k+1)^{-1}$. As $k$ grows, the
rate of these scheme, $R=k/(k+1)$ approaches $1$ and the
corresponding threshold tends to $0$. For large $k$, the scheme
tends to operate close to the capacity bound for high rates.

\begin{figure}[h]
\begin{center}
\includegraphics[width=0.85\columnwidth,draft=false]{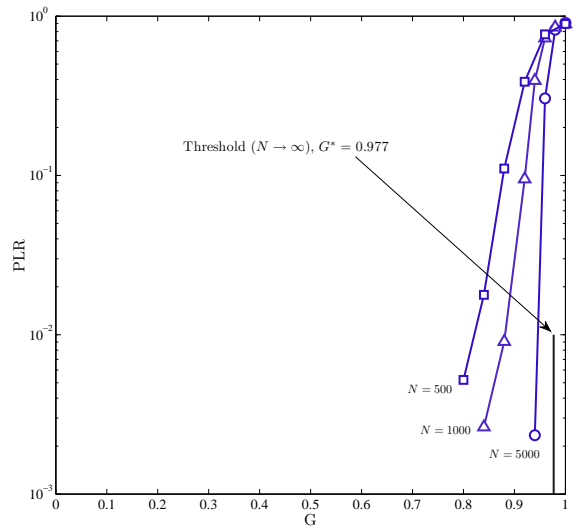}
\end{center}
\caption{PLR for the RA scheme based on the distribution $\Lambda_1(x)$. $N=5000, 1000,
500$, maximum iteration count set to $100$.}\label{fig:PLR}
\end{figure}

Fig.~\ref{fig:PLR} shows the packet loss rate (PLR) achieved by the
scheme employing the distribution $\Lambda_1(x)$. The results have
been derived via Monte Carlo simulations for MAC frames of size
$N=5000$, $1000$ and $500$ slots, and are compared with the capacity
of the scheme, $G^*=0.977$. For the $N=5000$ case, a PLR close to
$2\cdot 10^{-3}$ is achieved at a channel traffic
$G=0.94\,\mathrm{[packets/slot]}$, only
$0.05\,\mathrm{[packets/slot]}$ away from the bound established by
Theorem~\ref{theorem:capacity_bound}
($\simeq0.99\,\mathrm{[packets/slot]}$).

\section{Conclusions}\label{sec:conclusions}
We introduced a RA protocol for the CCw/oFB which achieves large
efficiencies with mild coordination demands. The scheme is based on
the use to erasure correcting codes to recover packet segments that
are lost in collisions, and on SIC to resolve collisions. The
proposed protocol achieves reliable communication in the asymptotic
setting and attains capacities close to $1\,\mathrm{[packet/slot]}$.
A simple capacity bound as a function of the code rates employed to
encode the segments has been proved. The derived bound can be
approached by means of judiciously designed component code
distributions.



\end{document}